# Sub-10 nm Probing of Ferroelectricity in Heterogeneous Materials by Machine Learning Enabled Contact Kelvin Probe Force Microscopy

*Sebastian W. Schmitt[1,*], Rama K. Vasudevan[2], Maurice Seifert[1], Albina Y. Borisevich[2], Veeresh Deshpande[1], Sergei V. Kalinin[2] and Catherine Dubourdieu[1,3,*]*

[1] Helmholtz-Zentrum Berlin für Materialien und Energie, Functional Oxides for Energy Efficient Information Technology, Hahn-Meitner Platz 1, 14109 Berlin, Germany

[2] Oak Ridge National Laboratory, Center for Nanophase Materials Sciences, 1 Bethel Valley Rd., Oak Ridge, TN 37831-6487, USA

[3] Freie Universität Berlin, Physical Chemistry, Arnimallee 22, 14195 Berlin, Germany

* sebastian.schmitt@helmholtz-berlin.de, catherine.dubourdieu@helmholtz-berlin.de

ABSTRACT

Reducing the dimensions of ferroelectric materials down to the nanoscale has strong implications on the ferroelectric polarization pattern and on the ability to switch the polarization. As the size of ferroelectric domains shrinks to nanometer scale, the heterogeneity of the polarization pattern becomes increasingly pronounced, enabling a large variety of possible polar textures in nanocrystalline and nanocomposite materials. Critical to the understanding of fundamental physics of such materials and hence their applications in electronic nanodevices, is the ability to investigate their ferroelectric polarization at the nanoscale in a non-destructive way. We show that contact Kelvin probe force microscopy (cKPFM) combined with a k-means response clustering algorithm enables to measure the ferroelectric response at a mapping resolution of 8 nm. In a $BaTiO_3$ thin film on silicon composed of tetragonal and hexagonal nanocrystals, we determine a nanoscale lateral distribution of discrete ferroelectric response clusters, fully consistent with the nanostructure determined by transmission electron microscopy. Moreover, we apply this data clustering method to the cKPFM responses measured at different temperatures, which allows us to follow the corresponding change in polarization pattern as the Curie temperature is approached and across the phase transition. This work opens up perspectives for mapping complex ferroelectric polarization textures such as curled/swirled polar textures that can be stabilized in epitaxial heterostructures and more generally mapping the polar domain distribution of any spatially-highly-heterogeneous ferroelectric materials.

Nanoscale ferroelectrics offer great potential for applications in information technologies and energy saving devices.[1] Shrinking down the thickness and the lateral dimensions has strong implications on the ferroelectric polarization pattern and on the ability to switch the polarization.

At reduced size and under confinement, stress, electrical boundary conditions and surface-related effects are key to the stabilization of polarization textures and to the ability to screen charges upon switching.[2–5] In these conditions, curled polarization and non-Ising domain walls may emerge, which may result in polarization textures with lateral dimensions of 10-20 nm such as vortices or skyrmions[6,7] and their disordered analogs. Another effect of reduced size and stress is the possible stabilization of metastable phases, an effect that is used e.g. for the stabilization of high symmetry phases in polycrystalline $HfO_2$ films for boosting its dielectric permittivity or inducing ferroelectricity.[8,9] Polycrystalline $HfO_2$-based ferroelectric thin films are typically highly heterogeneous materials with a mixture of the ferroelectric orthorhombic $Pca2_1$ phase together with non-ferroelectric monoclinic or tetragonal phases and with grain sizes of 5-20 nm.[10–12] Advanced transmission electron microscopy methods are key in evidencing the crystalline phases in thin heterogeneous films and the curled polar textures.[6,7,10,13,14] Besides these methods, it is desirable to have other means to probe in a more versatile and nondestructive way the nanoscale pattern of the ferroelectric polarization. Piezo-response force microscopy (PFM) is a powerful non-destructive method for the investigation of nanoscale piezo- and ferroelectric phenomena.[15–18] In PFM the response amplitude is a measure of the effective piezoelectric coefficient $d_{zz}$, which can be related to the polarization magnitude, while the polarization direction can be determined from the phase signal. Via the detection of PFM responses at surfaces of various ferroelectrics, electromechanical response polarization domain imaging in a sub-10-nm resolution has been achieved.[19–21] However, since electrostatic effects and ionic motion can lead to the same typical PFM amplitude and phase responses as those of a ferroelectric material, PFM does not permit to unambiguously proof ferroelectricity.[18,22,23] The introduction of contact Kelvin probe force microscopy (cKPFM), a PFM-related method that usually is performed in combination with spectroscopic band excitation (BE) of the sample, has contributed to solve this issue by revealing a more detailed voltage-dependence of the measured electromechanical signal.[24,25] With this method, ferroelectricity in ultrathin epitaxial $BaTiO_3$ films down to 1.6 nm on Si has been evidenced.[26] Analysis of BE spectroscopic PFM and cKPFM imaging data has been developed and provides a versatile tool kit for noise reduction, resolution enhancement or for the identification of material phase transitions and complex physical correlations based on experimental data.[27–32]

In this study we address the question of the ferroelectricity polarization pattern determination at a scale of less than 10 nm. We investigate a $BaTiO_3$ thin film that consists of a mixture of ferroelectric tetragonal and non-ferroelectric hexagonal phases, with a grain size in the range of five to tens of nanometers as determined by transmission electron microscopy. We show that cKPFM, synergistically enhanced by a k-means response clustering algorithm, permits to determine the polarization pattern with a discretized mapping of the electromechanical response at a resolution below 10 nm. The method can also be applied to temperature-dependent cKPFM, which paves the way towards imaging nanoscale ferroelectric textures and their evolution with temperature.

## Results and Discussion

### Nanocrystalline composite $BaTiO_3$ film on silicon

A nanocrystalline $BaTiO_3$ thin film (thickness 18 nm) was obtained by growing amorphous $BaTiO_3$ on an amorphous $SrTiO_3$-buffered single crystalline Si (001) substrate by molecular beam epitaxy and performing a subsequent crystallization by *in situ* thermal annealing at 500 °C (for the details see methods section and supplementary Fig. S1).[33,34] After annealing, the film consists of $BaTiO_3$ nanocrystallites on top of a Sr-silicate interface with Si(100). The lateral grain sizes range from ~ 5 to 20 nm in diameter, as shown by scanning transmission electron microscopy (STEM) images (Fig. 1a, additional TEM images can be found in the supplementary Fig. S2). In order to get information on the crystalline phase(s), a sliding fast Fourier transformation of fourteen TEM images, and a subsequent linear unmixing of the images in frequency space by a non-negative matrix factorization (NMF) algorithm was performed resulting in a set of six endmembers[35,36] (for details see methods section). Four of these endmembers are noisy and are attributed to regions of poor crystalline ordering. Two endmembers are found to clearly represent a tetragonal and a hexagonal pattern of $BaTiO_3$ (Fig. 1b) showing that tetragonal and hexagonal nanocrystals are coexisting in the films (we could not determine if they had a preferred orientation or texture). This coexistence is further supported by UV-Raman spectroscopy. Figure 1c displays the corresponding Raman spectrum together with the one measured on a $BaTiO_3$ poly-domain single crystal. Next to the Raman peaks characteristic of the $BaTiO_3$ tetragonal phase (indicated as T in Fig. 1c), we clearly observe a Raman peak at 635cm$^{-1}$ which is assigned to the $A_{1g}$ mode of the $BaTiO_3$ hexagonal phase (h-$BaTiO_3$) as previously reported for h-$BaTiO_3$ single crystals at 636 cm$^{-1}$.[37] The hexagonal phase of $BaTiO_3$[38] is a stable polymorph of $BaTiO_3$ besides the perovskite one, that can be synthesized as single crystals.[39] At room temperature, $BaTiO_3$ crystalizes in a ferroelectric tetragonal phase (t-phase), and transform to a cubic phase at ~ 120°C. The transition from the perovskite cubic to the hexagonal form occurs at a temperature of ~ 1450°C (see[40] and reference therein). It was shown that polycrystalline h-$BaTiO_3$ can be stabilized at temperatures below 700°C[41] in microcrystalline samples which can be attributed to surface energy similarly to the well-known stabilization of the high temperature tetragonal $ZrO_2$ phase at a much lower temperature than the one required for bulk ceramics.[42] Recently, a stress-induced cubic-to-hexagonal phase transformation was reported in $BaTiO_3$ thin films submitted to a bending stress and a temperature of 575°C.[43] Once cooled down to room temperature, tetragonal and hexagonal phases were found to coexist.[43] It was reported that the larger the tensile stress was, the larger the hexagonal phase volume fraction was.[43] In our stack deposited on Si, the thermal tensile stress while the sample is annealed at 500°C and then cooled down to room-temperature may be the driving force for the stabilization of the hexagonal phase. Ferroelectricity was evidenced in h-$BaTiO_3$ single crystals with a critical temperature of ~74 K (at 5 K, $P_s$ = 2 µC/cm$^2$, one order of magnitude lower than the polarization of t-$BaTiO_3$ at room temperature).[44] Hence, it is not ferroelectric at room temperature, nor does it exhibit local polar regions.[45] Regarding the tetragonal phase, one could speculate that the tensile strain would impart a full *a*-axis orientation. However, the Raman spectrum shows the presence of a peak at ~723 cm$^{-1}$ which is absent in purely *a*-axis single crystals and characteristics for the *c*-axis orientation.[46] The Raman spectrum of the tetragonal nanocrystalline grains is similar to the one obtained for polycrystalline tetragonal $BaTiO_3$ measured in the same polarization configuration.[47] The strong broadening of the peaks (see e.g. peaks at ~ 520 cm$^{-1}$ and ~720 cm$^{-1}$) originate from the nanometer-scale size of the grains (~ 5-20 nm from the TEM). Note that since the incident laser spot has a diameter of about 1 µm, the Raman measurements cannot provide information about the lateral distribution of the different $BaTiO_3$ crystal phases (tetragonal, hexagonal) in the film.

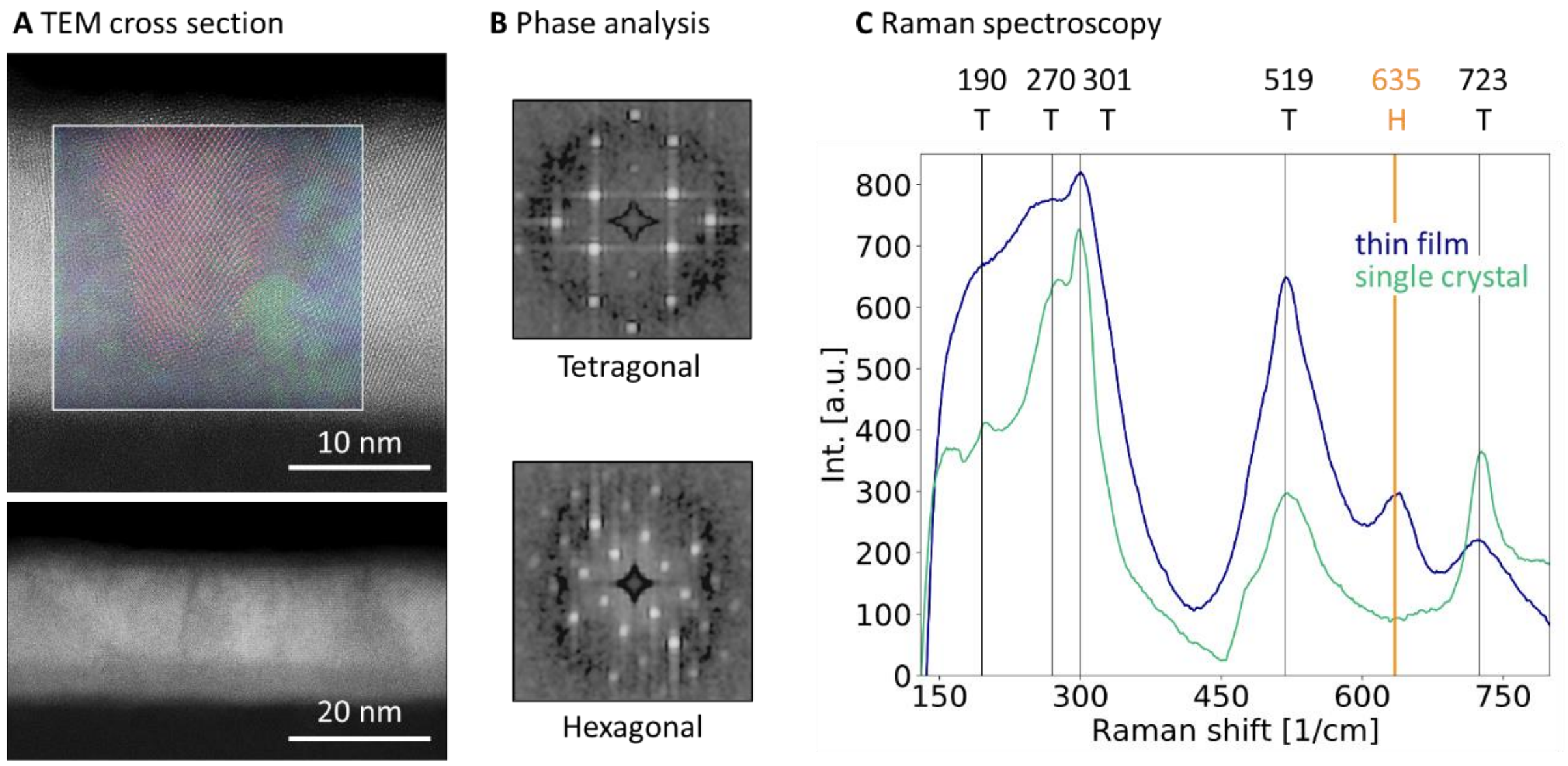

**Figure 1. (A) Two STEM images recorded in high angle annular dark field mode. In the white rectangle (top image) regions with distinct diffractogram patterns observed are shaded in red, green and blue colors. (B) Phase analysis by sliding FFT and NMF unmixing perfomed on the TEM cross section images. The greyscale images are two of the six endmembers of the NMF unmixing representing the tetragonal and hexagonal patterns in frequency space. (C) Raman spectra (in the $Z(XY)\overline{Z}$ polarization configuration) of the $BaTiO_3$ thin film (blue line) and of a $BaTiO_3$ poly-domain single crystal (green line). Indicated Raman peaks labeled with T (black) and H (orange) correspond to $BaTiO_3$ tetragonal and hexagonal phase, respectively.**

## Spectroscopic cKPFM measurements at the nanoscale

The electromechanical response of the samples was investigated using contact Kelvin probe force microscopy (cKPFM) with a DC write voltage from -7 to 7 V and a DC read voltage from – 6 to 8 V (for details see Methods section). The DC read voltage is shifted by +1V compared to the DC write voltage to account for the imprint that is observed (shift of the polarization loops towards positive voltage). In the applied cKPFM configuration the normal cantilever deflection is recorded, hence the out-of-plane electromechanical response (along the surface normal) is measured. Figure 2a displays the cKPFM responses recorded in 8 different positions along the X-direction measured on the same sample as the one analyzed in Fig. 1. The measurements shown in the upper rows are each 80 nm apart from the previous one while in the lower row, they are 8 nm apart. Figure 2b shows that significant qualitative as well as quantitative differences are observed in the cKPFM response data even for those recorded at locations that are only 8 nm apart from each other. The absence of hysteresis loop (quasi-linear response) shows the absence of out-of-plane ferroelectric polarization in these positions.[24] At other locations, hysteresis loops are observed. The shape of the curves indicates potential electrostatic and electrostrictive[48] signal contributions in addition to the ferroelectric contribution. In our data we can see only very small shifts of the surface potential from the write bias pulses. This does not affect the interpretation of the measurement, which focuses on the nonlinear aspects of the response as a function of read bias, indicating the presence of switchable polarization.[24] The temperature dependence of the cKPFM response discussed later clearly shows that the hysteretic behavior disappears which indicates that ferroelectricity is the predominant contribution. The amplitude of the hysteresis loops along the Y-axis for $V_{write}$ of -7 or 7 V varies with the location on the surface, which indicates different strengths of the out-of-plane ferroelectric polarization. These measurements indicate a heterogeneous distribution of polarization at a spatial scale down to 8 nm and we now further investigate how to extract information from mappings performed with such a low spatial scanning step.

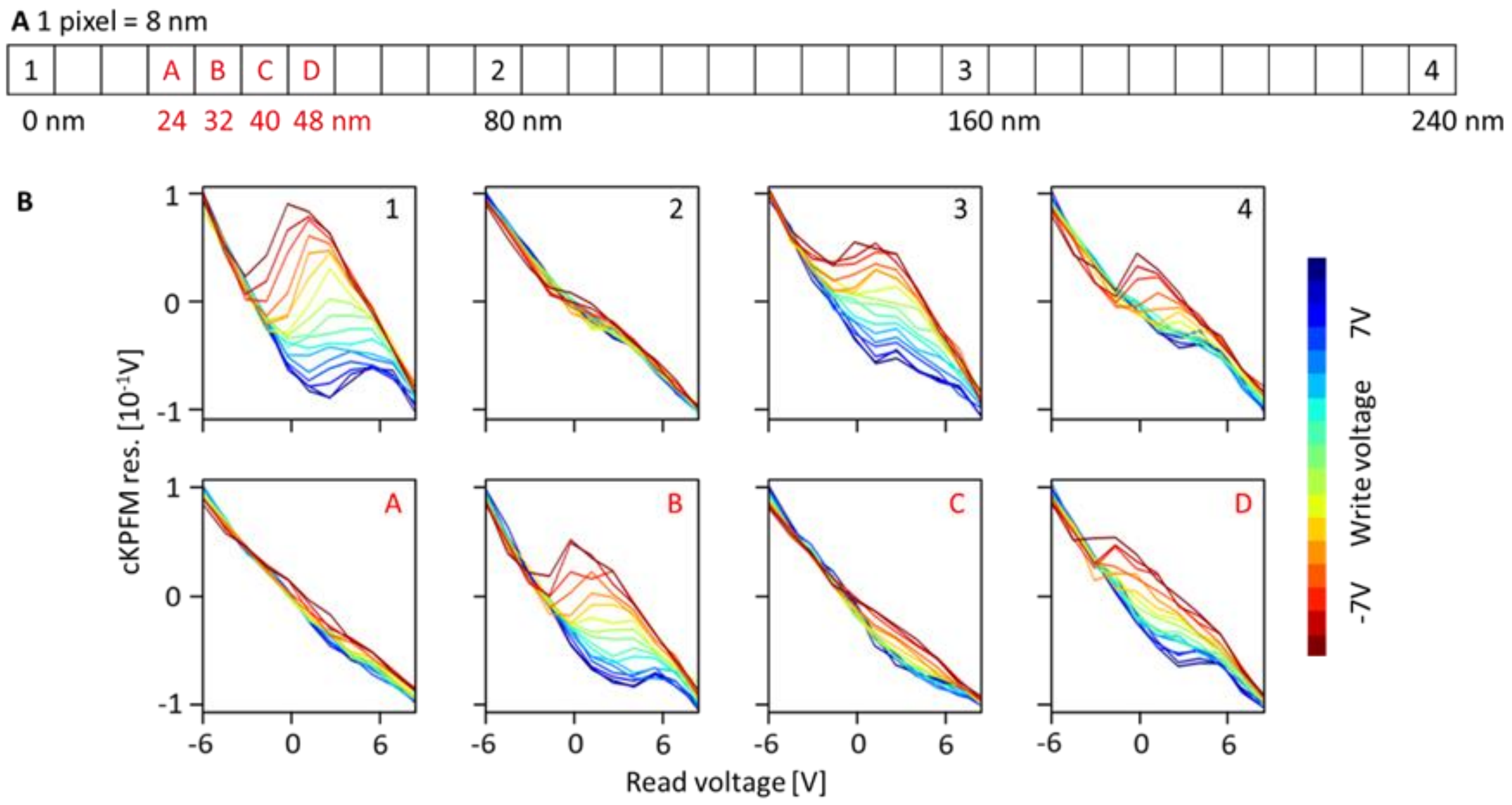

**Figure 2. (A) Schematical positions of the 8 cKPFM measurements (1, 2, 3, 4, and A, B, C, D). (B) cKPFM measurements performed at 8 different locations with a write voltage $V_{write}$ from -7 to 7 V and a read voltage $V_{dc}$ from – 6 to 8 V. The graphs shown in the upper row show measurements each 80nm apart from the previous one while in the lower row graphs corresponds to measurements each 8nm apart from the previous one.**

## Mapping the polarization combining cKPFM sub-10nm probing and machine learning analysis

Several cKPFM mappings were recorded with a pixel size of 8 x 8 $nm^2$ over different areas of 96 x 96 $nm^2$ (12 x 12 = 144 points map) up to 400 x 400 $nm^2$ (50 x 50 = 2500 points map) on the sample described in Fig. 1. The AFM topography scan shown in Fig. 3a was performed after a cKPFM mapping of 2500 points performed in the white rectangle region of this same image, which took about 8 h (11.3 s/point - 1408 measurements per point / pixel, see methods section). Comparing the topography of this region to the one of the surrounding sample surface, it is shown that the cKPFM mapping does not induce modifications of the surface. The noise on each cKPFM responses (such as those presented in Fig. 2b) is critically affected by local surface conditions, the indentation force, and the local tip-sample interaction. To reliably evidence ferroelectricity and discriminate between points of different ferroelectric response at the scale of few nanometers, an appropriate cKPFM signal processing has to cope with response fluctuations caused by e.g. local variabilities related to the local tip contact conditions. In particular, this applies for cKPFM signals on a lateral scale of 8 nm (see lower row in Fig. 2b) well below the radius of the applied AFM probes (~25 nm). To achieve this goal, we use a k-means clustering algorithm, to average over lateral fluctuations of the cKPFM responses. The k-means clustering is a method of vector quantization to partition a set of n observations into k clusters by minimizing the variances (Euclidean squared distances) within the clusters.[49] Each observation belongs to the cluster with the nearest mean which is called the cluster centroid. The n observations can then be represented by k cluster centroids which preserve the data characteristics and suppress the noise specific to an individual observation. The ideal number k of clearly discernible clusters which do not overdetermine n responses can be found by the so-called elbow method.[50] It uses the sum of variances between all n responses and their respective assigned cluster centroid and determines the number k in a way that adding another cluster does not significantly reduce the sum of variances any further. Mathematically this is the point of the highest curvature in the plot of the sum of variances as a function of k.[50] In this study, we find an optimum cluster number k=3 for all investigated regions of the sample (supplementary Fig. S3). The optimum k value is confirmed by dendrograms as illustrated in Fig. 3b showing the first three hierarchical levels of an agglomerative cluster analysis of the 2500 measurements.[51,52] It can be seen that a subdivision of the 2500 measurements (along the horizontal axis) in more than k=3 clusters results in only little additional cluster separation (vertical axis). Figure 3c shows the corresponding 3 centroids (C1, C2, C3). As expected, the centroids exhibit a reduced noise as compared to the individual measurements. Of the three clusters, one (C1) represents no ferroelectricity and the two others (C2, C3) represent two distinct out-of-plane ferroelectric polarization responses. The C1 response is assigned to the hexagonal phase, which is nonpolar at room temperature.[44] The centroids of clusters C2 and C3 are associated to the nanosized tetragonal $BaTiO_3$ crystallites. The two centroids reflect that the polarization amplitude is not unique which we attribute to the different nanocrystallite orientations and spread of their size (typically 5-20 nm lateral size from the TEM images). The resulting cKPFM map in terms of three cluster centroids map is shown in Fig. 3d. The spatial arrangement of clusters 1, 2 and 3 in the mapping, which occur at a relative abundance of 0.28, 0.45 and 0.27, illustrates the distribution of the out-of-plane ferroelectric polarization in high resolution on a 400 × 400 $nm^2$ area (another cKPFM mapping with 8 x 8 $nm^2$ pixel size, 2500 points map, is shown in supplementary Fig. S4 and indicates similar abundances of the three clusters). The different mappings we have performed show that partitioning of the cKPFM response data in clusters of discernible ferroelectric response clearly enhances the ability to image the ferroelectric polarization distribution in heterogeneous materials with nanoscale resolution, which is well below the resolution limit set by the size of the AFM tip.

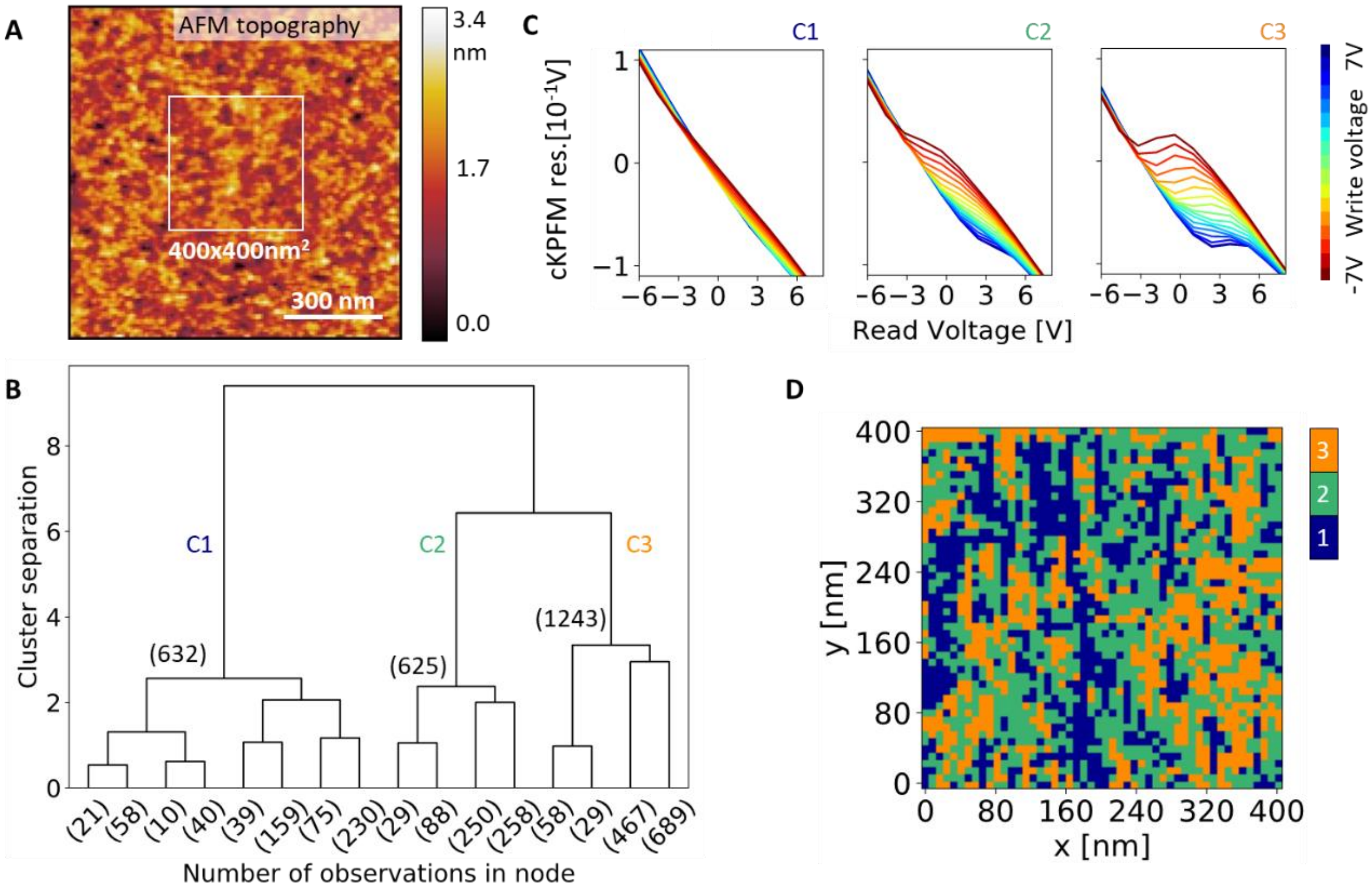

**Figure 3. (A) Contact AFM topography on the $BaTiO_3$ thin film sample shown in Fig 1. The white rectangle of 400 x 400 $nm^2$ inside the image indicates the region where the 2500 points cKPFM mapping with 8 x 8 $nm^2$ pixel size was performed (50 x 50 points map) (B) Dendrogram showing the first three hierarchical levels of an agglomerative cluster analysis of the 2500 measurements. Red numbers indicate the number of observations in the three main nodes, corresponding to the three clusters C1, C2, and C3. (C) The three centroids C1, C2 and C3 resulting from the k-means cluster analysis of the 2500 measurements (k=3). (D) cKPFM map with an area of 400 x 400 $nm^2$ and 8 x 8 $nm^2$ pixel size (50 x 50 = 2500 points map) displayed in terms of three cluster centroids C1, C2, and C3.**

## Repeatability

In order to check the consistency of such analysis, two subsequent cKPFM mappings with 8 x 8 $nm^2$ pixel size on the same 96 x 96 $nm^2$ area of the sample (12 x 12=144 points map) were performed. The k-means algorithm analysis results again in an optimum value of k=3. In Fig. 4, the resulting cKPFM maps are displayed in terms of these three centroids. The two maps exhibit a high coincidence, which is due to the unambiguous assignment of the measurements to one of the clearly distinct data clusters. The discrepancies are due to response intensities that can be assigned to neighboring clusters (C1/C2 or C2/C3). One artefact (C1/C3 assignment) observed arises at the corner of a map (typically 2 first pixels) or at a side of a map. One reason could be that the tip - when brought back from a previous scan position - may have picked up a contaminant that is then removed after the subsequent contact measurements. When applying a data clustering with k=4 (not sufficiently distinct clusters), the coincidence of the mappings decreases significantly, as shown in Fig. 4b, confirming that this clustering is not appropriate.

These results show that even though the cKPFM mappings are recorded at 8 nm sampling which is about 3 times lower than the radius of the used AFM probe (~ 25 nm) the measurements of the local response are highly reproducible (see also one further repeated cKPFM mapping in supplementary Fig. S5) The good reproducibility of the method in high spatial resolution is a result of combining the cKPFM oversampling (8 nm step size with a tip of radius ~25 nm) with the k-means cluster analysis, which reduces systematic noise of the response signal in the individual pixel.

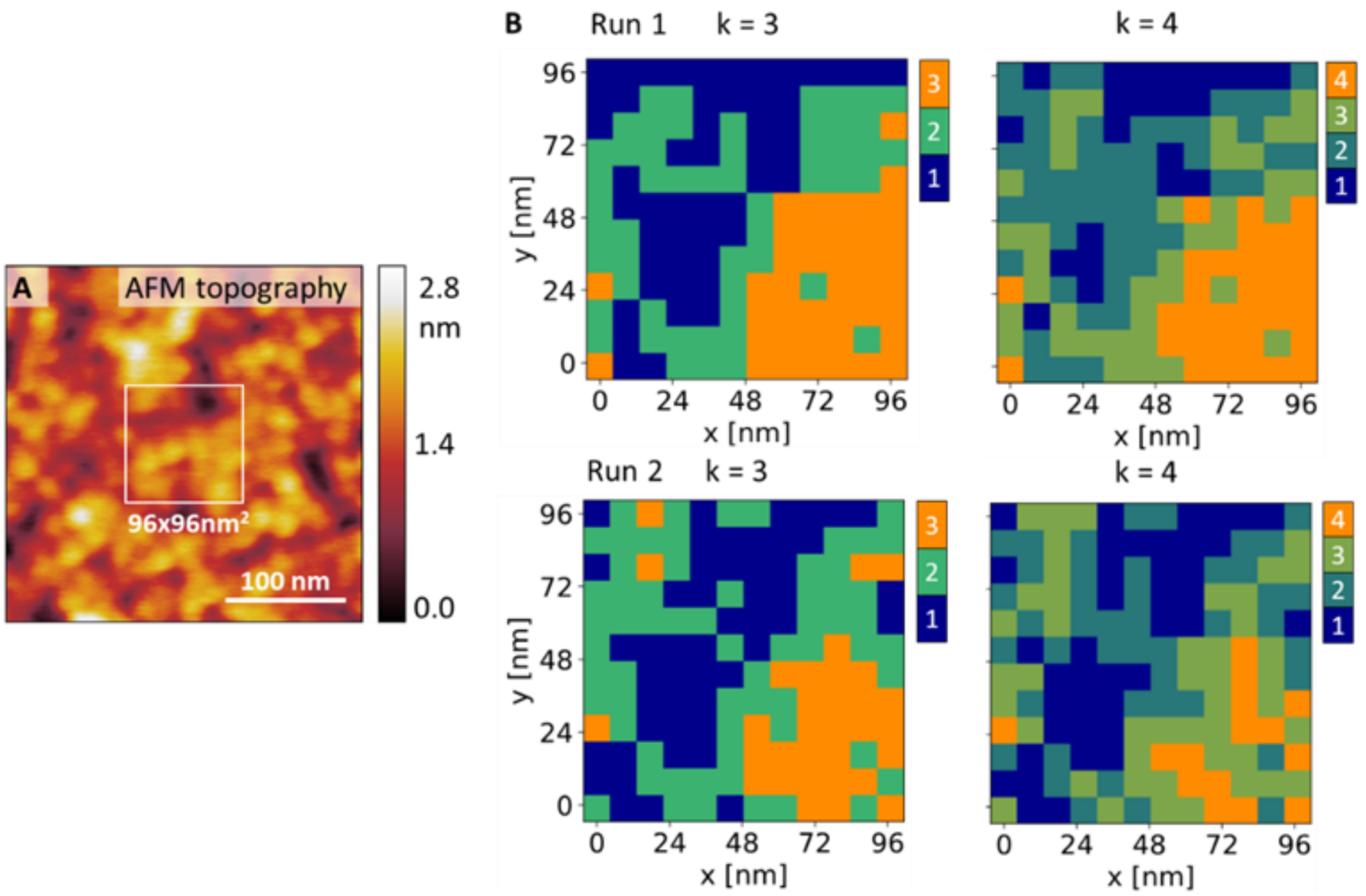

**Figure 4. (A) Contact AFM topography of a 300 x 300 nm$^2$ region of the $BaTiO_3$ thin film. The white rectangle indicates the positions of 96 x 96 nm$^2$ cKPFM mappings with 8 x 8 nm$^2$ pixel size (12 x 12 = 144 points map). (B) The cKPFM response in all 144 mapping points is displayed in terms of three cluster centroids of a k-means algorithm as in Fig. 3. Images show the results of two consecutive mapping runs (Run 1, Run 2) on the same location of the sample and for a clustering with k=3 or k=4.**

## Temperature evolution

In order to further test our method, cKPFM mappings were repeated at 7 temperatures of the sample scanning stage (25, 35, 45, 55, 65, 75, 90, and back to 25 °C). The scanned area was 3.2 x 3.2 μm$^2$, with a 400 x 400nm$^2$ pixel size (8 x 8 = 64 points map). Here the resolution was lowered in order to scan a large area and preserve the tip during the consecutive measurements (2 tips were used in total for the 8 consecutive mappings). The k-means clustering (k=3) was applied to the 64 data sets of each of the 8 measurements. The coordinates of the centroids were kept the same for the sets of data measured at different temperatures (no significant contribution from electrostatic charging was found in the measured loops that might gradually change once the temperature is rising). The cKPFM maps are shown in Fig. 5a. In Fig. 5b, the relative abundances of clusters C1, C2 and C3 are plotted as a function of the temperature. The abundance of measurements in cluster C1 is rising with increasing temperature from 25 to 90 °C, while concomitantly the number of responses in cluster C2 and C3 are decreasing and become eventually zero at 90 °C. For a temperature of 90 °C all measurements belong to cluster C1, corresponding to a weak or to no out-of-plane ferroelectric polarization in the $BaTiO_3$ thin film. After cooling the sample back to 25°C, the probed region is again ferroelectric (with 4 pixels showing no polarization). It is not possible to determine whether the exact same probe positions have been measured at the different temperatures since thermal drift produced a spatial shift that needed to be corrected after the adjustment of each temperature. From these experiments, the transition temperature from the tetragonal to the paraelectric cubic $BaTiO_3$ phase is between 75 °C and 90 °C. The lower Curie temperature as compared to the bulk one (~120 °C) is attributed to the reduced size of the crystallites. This observation is in good agreement with the temperature of 90 K corresponding to the maximum of the dielectric response obtained in $BaTiO_3$ ceramics with an average grain size of 8 nm[53] and with the transition temperature value of 80 °C reported for colloidally synthesized individual nanocrystals.[54] The detection of a phase transition temperature gives further evidence that the measured cKPFM response can directly be associated with the ferroelectricity of the investigated $BaTiO_3$ thin film and that the k-means clustering of a large set of data is a powerful method for the processing of cKPFM mapping data.

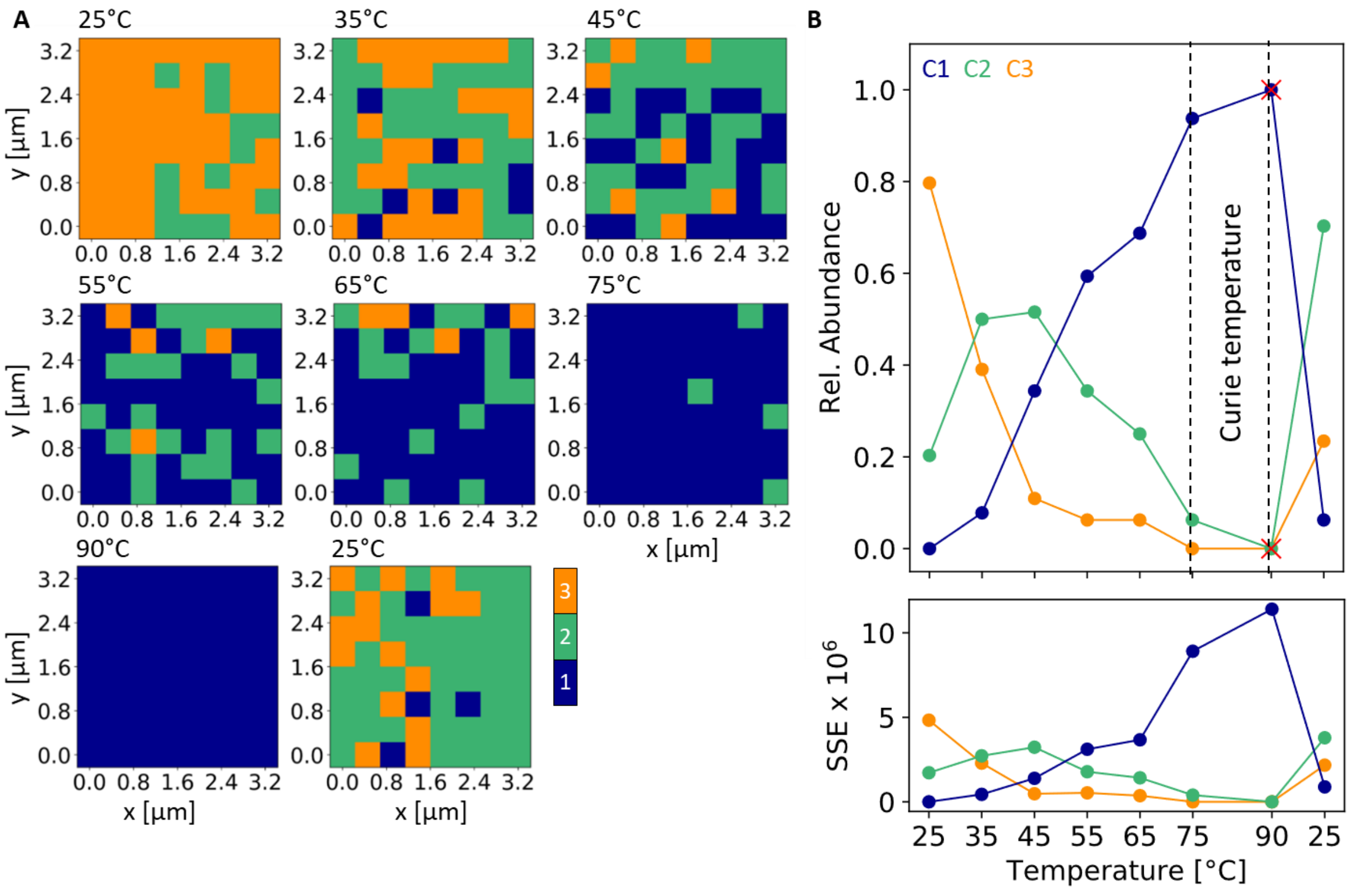

**Figure 5. (A) cKPFM maps for different temperatures displayed in terms of three cluster centroids of a k-means algorithm - on an area of 3.2 x 3.2 µm2 with 400x400 nm2 pixel size (8 x 8 points map). (B) Temperature evolution of the relative abundance of the three clusters in the mappings in Fig. 5a. The dashed black lines indicate the range in which the Curie temperature of the $BaTiO_3$ thin film is located. The red crosses indicate that the mapping at 90°C was performed in the same sample position after changing the AFM probe. The plot below shows the temperature evolution of the cluster-wise sum of squared errors (SSE).**

## Conclusion

We show that k-means clustering of cKPFM mapping data permits to discriminate signals associated to different ferroelectric responses with a high accuracy, repeatability, and high spatial resolution below 10 nm. This resolution is well below the resolution set by the AFM probe tip size. In a polycrystalline $BaTiO_3$ thin film on silicon in which h-$BaTiO_3$ and t-$BaTiO_3$ phases coexist, maps of cluster distribution of the ferroelectric polarization amplitude are obtained with an 8x8 $nm^2$ spatial resolution. This work opens up several perspectives for the study of ferroelectric materials in which the polarization changes at nanometer-size scale such as in ultrathin films or in nanostructures and could be used to probe swirling nanoscale polar textures. The method could also be used to map polar nanoregions in ferroelectric relaxor materials.[55] Moreover, as measurement mapping can be performed at different temperatures with a heating scanning stage, our method allows to follow the change in the distribution of the polarization amplitude with temperature and to determine the transition temperature of the ferroelectric phase to the paraelectric phase and more largely, any transitions that can occur in the materials with change in the ferroelectric polarization. Hence, the proposed method offers a promising pathway for the *in-situ* investigation of temperature-dependent polarization changes and pattern evolution in ferroelectrics at the nanoscale.

## Methods

**Sample fabrication.** Molecular beam epitaxy of amorphous 4 nm $SrTiO_3$ was performed on a Si(100) pre-structured substrate. The substrate consisted of a 300 nm thick $SiO_2$ film on top of Si(100) etched out in square area. The area studied here were of 100 x 100 $\mu m^2$ or 75 x 75 $\mu m^2$. The pre-structured substrate was cleaned in HF-last solution to remove the native oxide on top of the Si area. As compared to the process used to grow epitaxial $SrTiO_3$ films on Si,[33,34] the first initial two monolayers of amorphous $SrTiO_3$ were not crystallized such that further $SrTiO_3$ (3.2 nm) and $BaTiO_3$ (18 nm) grown on top were amorphous. $BaTiO_3$ was grown at 500 °C under $1 \times 10^{-7}$ Torr $O_2$ pressure. The sample was *in situ* annealed after deposition for 40 min at 500 °C in an $O_2$ atmosphere ($1 \times 10^{-5}$ Torr). Since the deposited stack under study was enclosed in recessed areas of 100 x 100 μ$m^2$ or 75 x 75 μ$m^2$ (see supplementary Fig. S1) no X-ray diffraction could be performed.

**STEM measurements.** STEM cross section lamellae of the stack were prepared at the position of the AFM, cKPFM and Raman measurements using focused ion beam (FIB) milling (Hitachi NB500 dual beam FIB-SEM). STEM images were recorded using Nion UltraSTEM 200 STEM operated at 200kV.

**Raman measurements:** Raman measurements were performed in a Horiba LabRam Evolution Raman spectrometer with a laser wavelength of 325 nm and an excitation power of ~25 μW/μ$m^2$ (40x NUV objective). Spectra were collected in backscatter configuration along z direction. The polarization of the incident laser light was perpendicular to the polarization of the scattered light (both in plane of the film). This configuration is denoted as $Z(XY)\bar{Z}$. The scattered light was analyzed using an 1800 l/mm grating blazed at 400 nm and a silicon CCD detector with 1024 pixels.

**cKPFM measurements.** For the band-excitation cKPFM measurements an atomic force microscope (AFM) at Oak Ridge National Laboratory (Asylum Cypher) was used. Applied AFM probes with a Cr/Pt conductive coating were acquired from Budget Sensors (Multi75E-G, 75 kHz, 3 N/m). Measurements were performed in contact mode around a resonance frequency of 341 kHz with a bandwidth of 64 kHz, a measurement amplitude of 1V and a pulse duration of 8 ms. The superimposed band-excitation waveform consisted of a write pulse sequence in 32 steps between -7 and 7 V (0 V to 7 V, 7 V to -7 V, -7 V to 0 V / 1.125 V step width) in which each write pulse was followed by read pulse sequence between -6 V and 8 V in 11 steps (1.4 V step width). Due to polarization imprint we shifted the center of the read window by +1V to record the full loops. Both the DC and AC positive voltages are applied through the tip. The pulse sequence with 32 write and 11 read steps was repeated 4 times which accounts for a total cKPFM measurement duration of 8 ms x 32 x 11 x 4 = 11.3 s and a total of 1408 measurements per pixel. For temperature dependent measurements the AFM was equipped with a heating stage. Between measurements at each temperature step, a hold time of 30 min was introduced to establish temperature equilibration of the setup.

**cKPFM data treatment.** The 32 x 11 x 4 = 1408 measurements in every mapping pixel represent the response amplitude and phase over the 64 kHz frequency range repeated for 4 times for each of the 32 write and 11 read voltage steps. Response amplitude and phase were fitted using a single harmonic oscillator model. After correcting the phase offset for each measurement, the cKPFM response magnitude for each writing pulse sequence is plotted against the value of the read voltage resulting in the response spectra as shown in Fig. 2b. For each of the cKPFM mappings, a k-means cluster analysis (k = 2, 3, 4) over the response spectra in the individual grid points was performed, using the KMeans method of the Cluster module in the Python software library Scikit Learn (version 0.23.2). Mappings as shown in Figs. 3d, 4b and 5a were represented in terms of the cluster centroids calculated by the KMeans.labels method, the cluster centroids (as in Fig. 3c) were exported using the KMeans.cluster_centers method. The data formatting and representation was performed based on code from the Python software libraries pyUSID and Pycroscopy.[56] For the calculation and display of the hierarchical clustering dendrogram, the Dendrogram method from the Python Cluster module was used. To determine the point of the highest curvature in the plot of the sum of variances as a function of k (supplementary Fig. S3), we used the KneeLocator method from the Python library kneed.

## AUTHOR INFORMATION

**Corresponding Authors**

**Sebastian Schmitt** – Helmholtz-Zentrum Berlin für Materialien und Energie, 14109 Berlin, Germany; Email: sebastian.schmitt@helmholtz-berlin.de

**Catherine Dubourdieu** - Helmholtz-Zentrum Berlin für Materialien und Energie, 14109 Berlin, Germany and Freie Universität Berlin, 14195 Berlin, Germany; Email: catherine.dubourdieu@helmholtz-berlin.de

**Author Contributions**

C.D., S.W.S. and R.V. conceived the study. S.W.S. and R.V. performed the cKPFM measurements. S.W.S. performed the Raman measurements. A.Y.B. performed the TEM measurements. S.W.S., M.S. and R.V. performed the data analysis. C.D., S.W.S., R.V., and A.Y.B. discussed the data analysis. All co-authors discussed the results.  S.W.S. and C.D. wrote the manuscript. All co-authors read and edited the manuscript.

## ACKNOWLEDGMENT

C.D. acknowledges Martin Franck and Vijay Narayanan from IBM in Yorktown Height, NY, USA for providing the pre-structured Si wafer and Lucie Mazet for the growth of the sample at the technology platform “NanoLyon” of  INL / Ecole Centrale de Lyon, Ecully, France. C.D. and S.W.S. acknowledge access to the Center for Nanophase Materials Sciences through the proposals CNMS2018-298 and CNMS2020-B-00356 and thank Nina Balke for support and useful discussions.

SUPPORTING INFORMATION

The supporting information contains the supplementary Figures S1-S5.

# Supporting Information: Sub-10 nm Probing of Ferroelectricity in Heterogeneous Materials by Machine Learning Enabled Contact Kelvin Probe Force Microscopy

*Sebastian W. Schmitt[1,*], Rama K. Vasudevan[2], Maurice Seifert[1], Albina Y. Borisevich[2], Veeresh Deshpande[1], Sergei V. Kalinin[2] and Catherine Dubourdieu[1,3,*]*

[1] Helmholtz-Zentrum Berlin für Materialien und Energie, Functional Oxides for Energy Efficient Information Technology, Hahn-Meitner Platz 1, 14109 Berlin, Germany

[2] Oak Ridge National Laboratory, Center for Nanophase Materials Sciences, 1 Bethel Valley Rd., Oak Ridge, TN 37831-6487, USA

[3] Freie Universität Berlin, Physical Chemistry, Arnimallee 22, 14195 Berlin, Germany

* sebastian.schmitt@helmholtz-berlin.de, catherine.dubourdieu@helmholtz-berlin.de

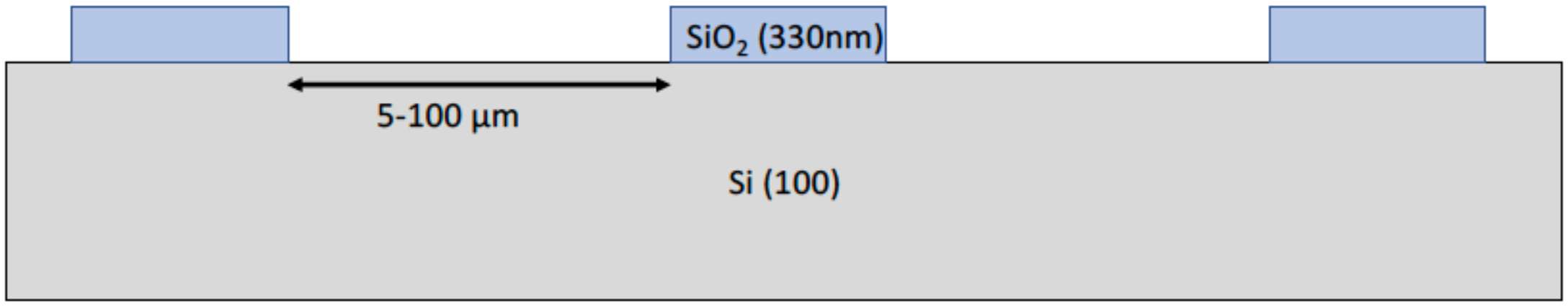

**Figure S1. The material investigated in this study was grown directly on the Si surface by MBE, in recessed areas of size of 100 x 100 µm$^2$ or below. Hence, no X-ray diffraction could be performed. The cKPFM measurements were performed in areas of 75 x 75 µm$^2$ and 100 x 100 µm$^2$ size**.

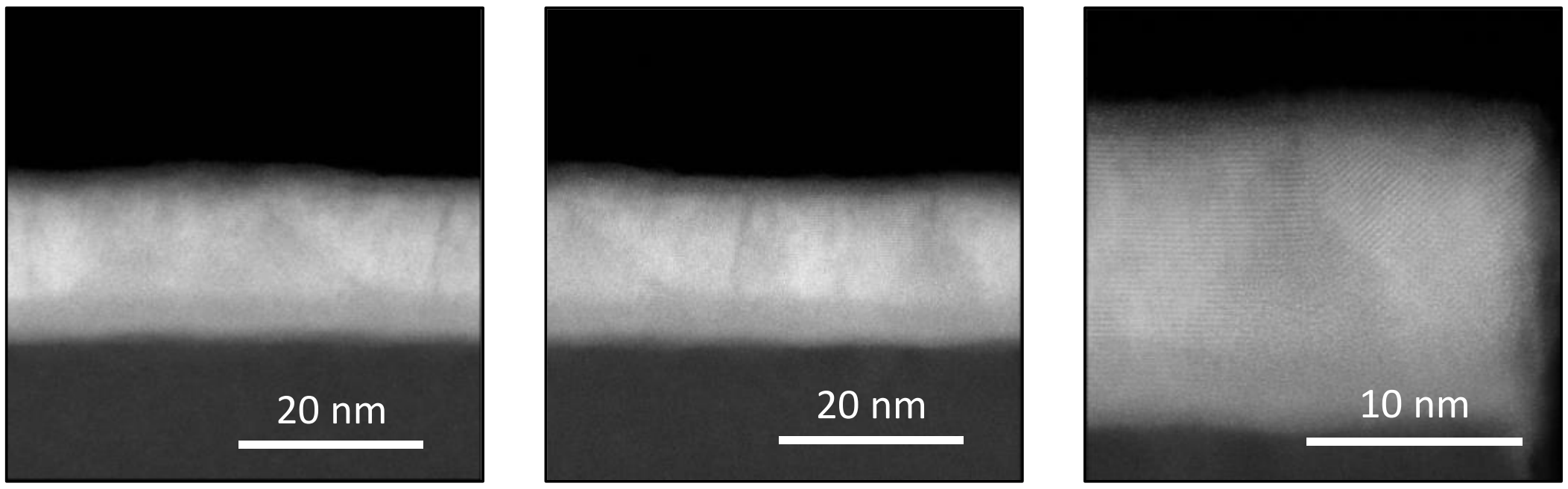

**Figure S2. Three STEM cross sections of the stack recorded in bright field mode.**

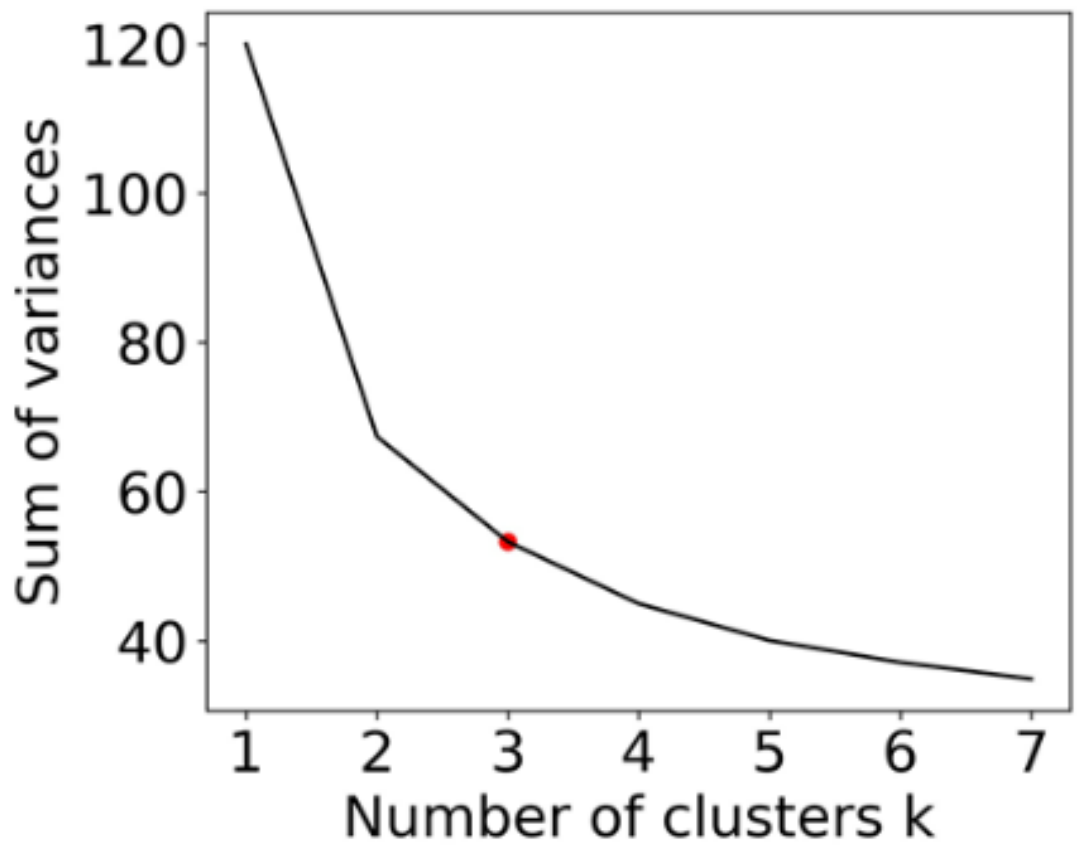

**Figure S3. Sum of variances between all n response observations and their respective assigned cluster centroid as function of cluster number k for the 400 x 400 nm$^2$ cKPFM map with 8 x 8 nm$^2$ pixel size (50 x 50 = 2500**

**points map, position indicated in Fig. 3a). The red point indicates the ideal number of clusters derived according to the elbow method (k=3).**

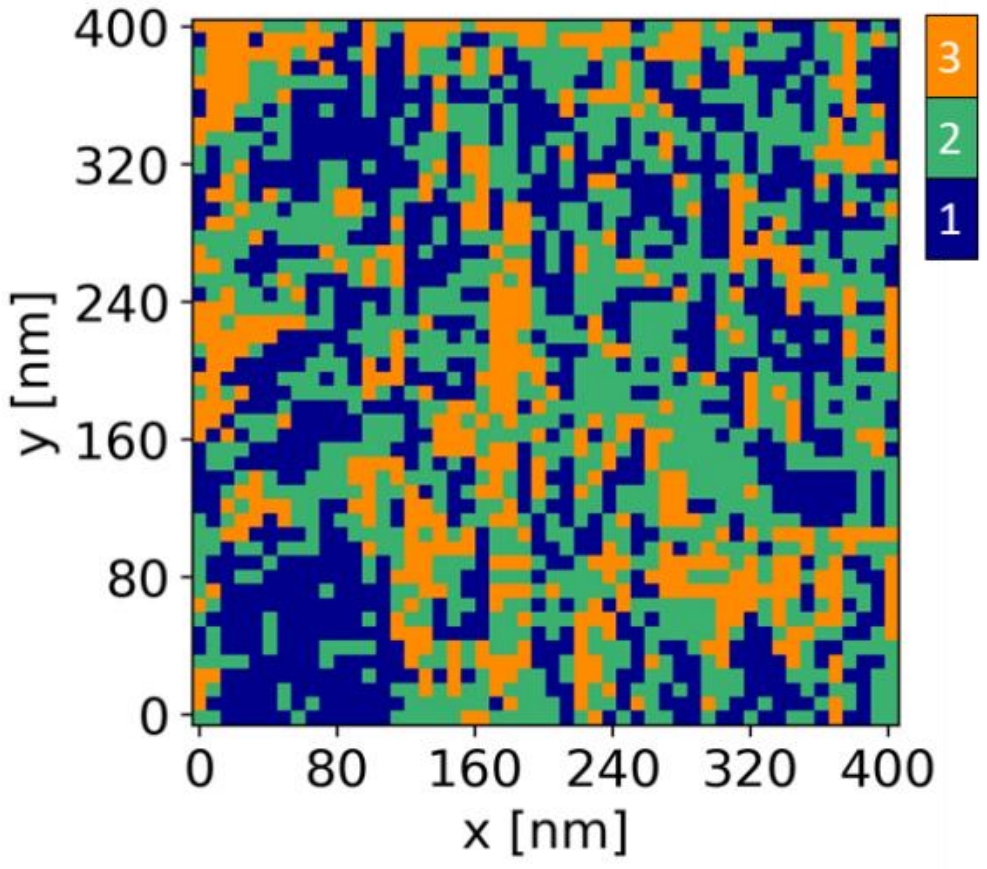

**Figure S4. cKPFM map with an area of 400 x 400 $nm^2$ and 8 x 8 $nm^2$ pixel size (50 x 50 = 2500 point map) displayed in terms of three cluster centroids C1, C2, and C3.**

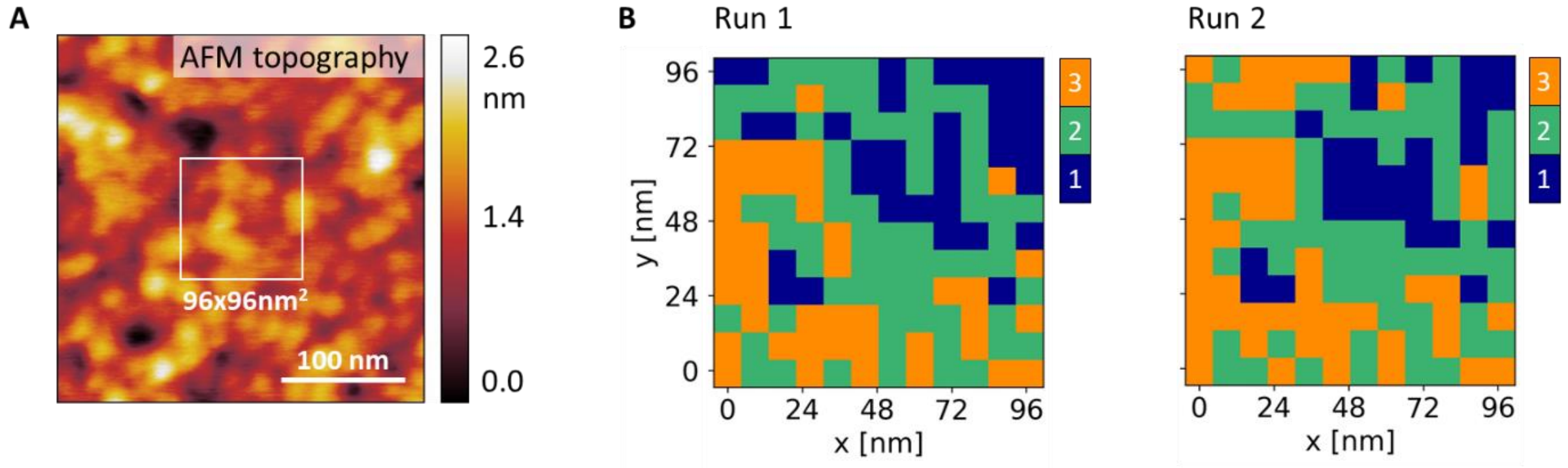

**Figure S5. (A) Contact AFM topograpy of a 300 x 300 $nm^2$ region of the $BaTiO_3$ thin film. The white rectangle indicates the positions of 96x96 $nm^2$ cKPFM mappings with 8 x 8 $nm^2$ pixel size (12 x 12 = 144 points map). (B) The cKPFM response in all 144 mapping points is displayed in terms of three cluster centroids of a k-means algorithm as in Fig. 3. Images show the results of two consecutive mapping runs (Run 1, Run 2) on the same location of the sample and for a clustering with k=3.**